\documentclass[letterpaper]{article} 
\usepackage{aaai2026}  
\usepackage{times}  
\usepackage{helvet}  
\usepackage{courier}  
\usepackage[hyphens]{url}  
\usepackage{graphicx} 
\usepackage{natbib}  
\usepackage{caption} 
\usepackage{algorithm}
\usepackage{algorithmic}

\usepackage{newfloat}
\usepackage{listings}
\DeclareCaptionStyle{ruled}{labelfont=normalfont,labelsep=colon,strut=off} 
\floatstyle{ruled}
\newfloat{listing}{tb}{lst}{}
\floatname{listing}{Listing}
\usepackage{amsfonts}
\usepackage{mathtools}
\usepackage{mathrsfs}
\usepackage{array}
\usepackage{booktabs}
\usepackage{makecell}
\usepackage{multirow}
\usepackage[table,xcdraw]{xcolor}
\usepackage{subcaption}
\usepackage{color}

\newcommand{\task}{\emph{Diffusion-Oriented Content Generation}}

\title{Designed to Spread: A Generative Approach to Enhance Information Diffusion}

\author{
    Ziqing Qian\equalcontrib\textsuperscript{\rm 1,\rm 2},
    Jiaying Lei\equalcontrib\textsuperscript{\rm 1,\rm 3}, 
    Shengqi Dang\textsuperscript{\rm 1,\rm 2,\rm 3},
    Nan Cao\textsuperscript{\rm 1,\rm 2,\rm 3}\thanks{Corresponding author.}
}
\affiliations{
    \textsuperscript{\rm 1}Intelligent Big Data Visualization Lab, Tongji University, Shanghai, China\\
    \textsuperscript{\rm 2}Shanghai Research Institute for Intelligent Autonomous System, Tongji University, Shanghai, China\\
    \textsuperscript{\rm 3}Shanghai Innovation Institute, Shanghai, China\\
    
    2411920@tongji.edu.cn,
    jiaying.lei@outlook.com,
    dangsq123@tongji.edu.cn,
    nan.cao@gmail.com
}

\begin{document}

\maketitle

\begin{abstract}
Social media has fundamentally transformed how people access information and form social connections, with content expression playing a critical role in driving information diffusion. While prior research has focused largely on network structures and tipping point identification, it provides limited tools for automatically generating content tailored for virality within a specific audience. To fill this gap, we propose the novel task of \task~(DOCG) and introduce an information enhancement algorithm for generating content optimized for diffusion. Our method includes an {influence indicator} that enables content-level diffusion assessment without requiring access to network topology, and an {information editor} that employs reinforcement learning to explore interpretable editing strategies. The editor leverages generative models to produce semantically faithful, audience-aware textual or visual content. Experiments on real-world social media datasets and user study demonstrate that our approach significantly improves diffusion effectiveness while preserving the core semantics of the original content.
\end{abstract}

\begin{links}
    \link{Code}{https://github.com/idvxlab/designed-to-spread}
\end{links}

\section{Introduction}

Social media has reshaped how people communicate, access information, and form connections. Platforms such as X, Facebook, and TikTok have become powerful channels for news dissemination, public discourse, and cultural influence. Information spreads via diffusion-like processes along user networks, motivating extensive research on diffusion mechanisms~\cite{bakshy2012role,guille2013information,weng2013virality}.

Extensive research has examined information diffusion in social networks, including modeling diffusion dynamics~\cite{granovetter1978threshold,pastor2001epidemic,kempe2003maximizing}, identifying influential seed users~\cite{chen2009efficient}, and limiting spread via edge removal~\cite{tong2012gelling}. These approaches typically depend on detailed network structure, which is often unavailable or incomplete in real-world settings. Meanwhile, how content is crafted also significantly affects its diffusion, especially for specific audiences. Yet, content-level optimization remains underexplored. This highlights the need for automated methods to generate audience-aware, diffusion-oriented content without relying on explicit network information.

The rapid development of AIGC techniques offers promising potential to fulfill the above needs. While these methods excel at generating high-quality text and image content~\cite{chen2023pixartalphafasttrainingdiffusion,podell2023sdxlimprovinglatentdiffusion}, optimizing content specifically for maximizing information diffusion remains challenging due to three key issues: (1) diffusion impact is difficult to measure at the content level, especially when network topology is unavailable or incomplete; (2) preserving the original message's intent while generating user-preferred representations is non-trivial, as misalignment may cause misunderstanding or even unintentional misinformation; and (3) designing a unified framework that integrates diffusion feedback into multimodal content generation remains an open challenge.

To address the above challenges, we introduce the task of \task~(DOCG), which aims to generate a semantically faithful variant (text or image) of a given message on a specific topic, optimized to maximize its diffusion impact among a target audience group. We propose an information enhancement framework with two key components: an \textit{influence indicator} and an \textit{information editor}. The influence indicator estimates content-level diffusion potential within the target audience. The information editor formulates content generation as a reinforcement learning (RL) problem, guiding a generative model to revise content through interpretable, modality-specific editing actions (e.g., amplifying emotional tone). The RL objective is to maximize predicted influence while preserving the original message’s semantics. Our main contributions are summarized as follows:

\begin{itemize}
\item We propose a new task, \task~(DOCG), which aims to generate audience-aware content to maximize diffusion influence.
\item We propose a reinforcement learning framework that reshapes input messages to enhance diffusion while preserving their original intent. It leverages an influence indicator to estimate content-level impact for target audiences without relying on network topology, using this feedback to guide iterative multimodal revisions.
\item We demonstrate the versatility of our method across both textual and visual modalities, achieving significant improvements on real-world social media dataset.
\end{itemize}

\section{Related Work}
In this section, we review related work in three areas most relevant to our study: social media analysis, information diffusion models, and content editing and generation.

\subsection{Social Media Analysis}

Social media analysis explores the complex interplay of factors that shape how information is created, interpreted, and propagated online. Early studies primarily adopted descriptive approaches to reveal observable trends~\cite{cheng2014can,de2021no,etta2023characterizing,flamino2023political}, drawing insights from three main perspectives. 

The first focuses on content virality. For example, Vosoughi et al.~\cite{vosoughi2018spread} found that false information spreads more rapidly than truthful content on X, largely due to its novelty and emotional appeal. The second examines user engagement behaviors, including posting frequency~\cite{benevenuto2009characterizing,spasojevic2015post}, reaction latency~\cite{hodas2014simple}, and cross-platform activity~\cite{xu2014quantifying,iamnitchi2023modeling,alipour2024cross}. Especially, Shahbaznezhad et al.~\cite{shahbaznezhad2021role} further demonstrated that content format and platform-specific algorithms significantly affect user responsiveness, with emotionally charged posts eliciting faster reactions. The third investigates the structural properties of diffusion cascades. Notarmuzi et al.~\cite{notarmuzi2022universality} identified universal propagation patterns in early retweet cascades, highlighting the role of network topology and user influence in determining final reach and virality.

While these studies offer valuable insights into diffusion dynamics, they remain largely observational and often overlook the causal impact of content transformation on user behavior. This gap motivates our generative approach to enhancing information diffusion through content design.

\subsection{Information Diffusion Model}

Research on information diffusion models seeks to model the content propagation through social networks. Early approaches were inspired by epidemiological processes, such as the Independent Cascade (IC) and Linear Threshold (LT) models; however, these models assume static networks and fixed influence probabilities among different people~\cite{goldenberg2001talk,kempe2003maximizing}.

To overcome these simplifications, subsequent work has incorporated temporal dynamics, deep learning, and evolving network structures.
For example, NetRate~\cite{gomez2012inferring} applies survival analysis to infer time-dependent transmission functions from cascade data. Inf-VAE~\cite{sankar2020inf} models latent diffusion representations via variational autoencoders, while DeepDiffuse~\cite{islam2018deepdiffuse} leverages deep neural networks to predict both the participants and timing of information cascades. More recently, Meng et al.~\cite{meng2025spreading} propose a data-driven model that captures the dynamics of social reinforcement and decay in online diffusion, offering a concise formulation for large-scale spread.

Despite these advancements, existing models still exhibit critical limitations: a strong reliance on explicit network structures, sensitivity to seed selection, and a lack of content awareness in modeling~\cite{jiang2023retweet}. These challenges underscore the need for a content-based diffusion influence indicator that functions independently of network topology.

\begin{figure*}[t]
    \centering
    \includegraphics[width=1.0\linewidth]{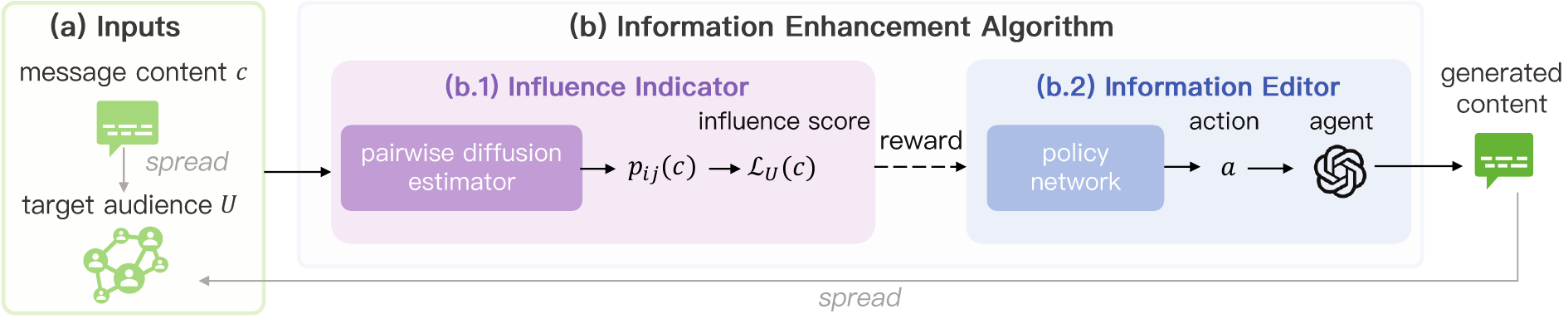}
    \caption{Overview of our proposed information enhancement framework, consisting of two main components: the {influence indicator}, and the {information editor}.}
    \label{fig:overview}
\end{figure*}

\subsection{Content Editing and Generation}

Content editing and generation synthesize new content aligned with a target distribution or specific objectives such as style transfer. Generation creates content from scratch, while editing modifies existing references to achieve desired transformations, preserving coherence and semantics.

Early work on content generation used variational autoencoders (VAEs)~\cite{kingma2013auto} and generative adversarial networks (GANs)~\cite{goodfellow2020generative} to learn latent representations for attribute-controlled editing~\cite{radford2016unsupervisedrepresentationlearningdeep,dumoulin2017adversariallylearnedinference}. With large-scale pretrained models, generation has become more fluent and controllable. Large language models (LLMs) such as GPT-3~\cite{brown2020language} and T5~\cite{raffel2020exploring} enable fine-grained text editing via prompts. In vision, diffusion models including Stable Diffusion~\cite{rombach2022high}, SDXL~\cite{podell2023sdxlimprovinglatentdiffusion}, and PixArt-$\alpha$~\cite{chen2023pixartalphafasttrainingdiffusion} support high-quality image generation and editing. Multimodal systems such as BLIP-Diffusion~\cite{li2023blip} and MIGE~\cite{Tian_2025} further unify the editing across modalities.

However, most methods focus on general-purpose generation. Although recent work explores engagement-driven generation for social networks~\cite{coppolillo2025engagement}, audience-specific preferences remain overlooked. Leveraging the adaptability of LLMs and diffusion models, we explore audience-aware content generation to better support information diffusion.

\section{Diffusion-Oriented Content Generation}
In this paper, we introduce a new task, \task~(DOCG), which aims to generate a semantically faithful variant (text or image) of a given message
$c$, optimized to maximize its diffusion impact among a group of target audiences $U$. We also propose a novel modality-free algorithm framework based on reinforcement learning to accomplish this task. 
The core idea of our framework is to evaluate the diffusion impact and select strategies to edit the original message iteratively based on pretrained generative models such as GPT-4 (for text) or Pixart-$\alpha$~\cite{chen2023pixartalphafasttrainingdiffusion} (for image). By repeating this evaluation–editing loop, the framework progressively enhances the expected diffusion impact, ultimately producing an optimized variant tailored for the group of target audiences. 

As shown in Figure~\ref{fig:overview}, our framework accepts an initial message $c$ and a group of target audiences $U$ as the inputs, and comprises two components: (b.1) the \emph{influence indicator} and (b.2) the \emph{information editor}. The influence indicator computes an overall influence score \(\mathcal{L}_U(c)\) to measure the diffusion impact.
The information editor treats \(\mathcal{L}_U(c)\) as the reward signal for a policy network that samples an editing action \(a\). This action is executed by a generative agent (e.g., a large language or text-to-image model) to produce a revised message. In the following sections, we describe the modeling and implementation details of each component.

\subsection{Influence Indicator}

The influence indicator produces an influence score to quantify the diffusion impact of the message content \(c\) within the group of target audiences \(U\). Formally, the influence score is defined as:
\begin{equation}\label{eq:L_def}
  \mathcal{L}_U(c)
  \;=\;\max_{i}\Big(\frac{1}{|U|} \sum_{u_j \in U} p_{ij}(c)\Big)
\end{equation}
where \(p_{ij}(c)\) denotes the probability that user \(u_j \in U\) will spread (e.g., retweet or reply to) content \(c\) after being exposed to it from user \(u_i \in U\). This formulation captures the diffusion potential of the most influential initiator.
In the following, we detail the pairwise diffusion estimator for computing \(p_{ij}(c)\), its training data, and its loss function.

\underline{\textit{Pairwise Diffusion Estimator.}}
The pairwise diffusion estimator is implemented as a neural network that predicts the probability \(p_{ij}(c)\). The model operates in three stages: feature extraction, projection, and interaction. First, we extract the features \(f_i\), \(f_j\), and \(f(c)\) corresponding to users \(u_i\), \(u_j\), and content \(c\), respectively. User features are computed as the average of features from previously created or shared content, while content features are obtained using a pretrained CLIP encoder~\cite{clip}, which aligns the text and image features in a shared space. To handle long-text inputs, we adopt Long-CLIP~\cite{long-clip}, a variant of CLIP designed to overcome token length limitations.
Then, all input features are projected into a latent space via a shared Multilayer Perceptron~(MLP), denoted as \(M_1\). The combined representation $h$ is obtained by concatenating the projected vectors:
\begin{align}
  h = M_1(f_i) \oplus M_1(f_j) \oplus M_1(f(c))
\end{align}
where \(\oplus\) denotes vector concatenation. Finally, $h$ is passed through another MLP, \(M_2\), which captures nonlinear interactions among the user and content features. A sigmoid function \(\sigma(\cdot)\) is applied to obtain the final probability:
\begin{align}
  p_{ij}(c) = \sigma\bigl(M_2(h)\bigr)
\end{align}

\begin{table*}[]
\setlength{\tabcolsep}{1mm}
{
\begin{tabular}{|ccl|}
\hline
\rowcolor[HTML]{FFFFFF} 
{\color[HTML]{000000} }                                                         & {\color[HTML]{000000} \textbf{Dimension}}            & \multicolumn{1}{c|}{\cellcolor[HTML]{FFFFFF}{\color[HTML]{000000} \textbf{Description}}}                                                \\ \hline
\cellcolor[HTML]{FFFFFF}{\color[HTML]{000000} }                                 & {\color[HTML]{000000} \textit{Social Currency}}      & {\color[HTML]{000000} whether the text enhances the sharer’s profile (e.g., appearing good, intelligent, funny).} \\
\rowcolor[HTML]{FFFFFF} 
\cellcolor[HTML]{FFFFFF}{\color[HTML]{000000} }                                 & {\color[HTML]{000000} \textit{Triggers}}             & {\color[HTML]{000000} whether the text is strongly connected to real-life scenarios, promoting frequent recall.}                         \\
\cellcolor[HTML]{FFFFFF}{\color[HTML]{000000} }                                 & {\color[HTML]{000000} \textit{Emotion}}              & {\color[HTML]{000000} whether the text evokes high-arousal emotions (e.g., excitement, surprise, delight).}                              \\
\rowcolor[HTML]{FFFFFF} 
\cellcolor[HTML]{FFFFFF}{\color[HTML]{000000} }                                 & {\color[HTML]{000000} \textit{Public}}               & {\color[HTML]{000000} whether the text format is easily shareable and visible.}                                                          \\
\cellcolor[HTML]{FFFFFF}{\color[HTML]{000000} }                                 & {\color[HTML]{000000} \textit{Practical Value}}      & {\color[HTML]{000000} whether the text offers concrete guidelines or useful information.}                                                \\
\rowcolor[HTML]{FFFFFF} 
\multirow{-6}{*}{\cellcolor[HTML]{FFFFFF}{\color[HTML]{000000} \textbf{Text}}}  & {\color[HTML]{000000} \textit{Stories}}              & {\color[HTML]{000000} whether the text includes a narrative framework or memorable elements.}                                            \\ \hline
\cellcolor[HTML]{FFFFFF}{\color[HTML]{000000} }                                 & {\color[HTML]{000000} \textit{Colorfulness}}         & {\color[HTML]{000000} whether the image is rich in color and has visual impact.}                                                         \\
\rowcolor[HTML]{FFFFFF} 
\cellcolor[HTML]{FFFFFF}{\color[HTML]{000000} }                                 & {\color[HTML]{000000} \textit{Human Scene}}          & {\color[HTML]{000000} whether the image includes scenes with people to enhance emotional resonance.}                                     \\
\cellcolor[HTML]{FFFFFF}{\color[HTML]{000000} }                                 & {\color[HTML]{000000} \textit{Emotion}}              & {\color[HTML]{000000} whether the image can evoke strong emotional responses (such as joy, shock, or being moved).}                      \\
\rowcolor[HTML]{FFFFFF} 
\cellcolor[HTML]{FFFFFF}{\color[HTML]{000000} }                                 & {\color[HTML]{000000} \textit{Professional}}         & {\color[HTML]{000000} whether the image has the quality of professional photography.}                                                    \\
\cellcolor[HTML]{FFFFFF}{\color[HTML]{000000} }                                 & {\color[HTML]{000000} \textit{Brightness}}           & {\color[HTML]{000000} whether the image is bright and attention-grabbing.}                                                               \\
\rowcolor[HTML]{FFFFFF} 
\cellcolor[HTML]{FFFFFF}{\color[HTML]{000000} }                                 & {\color[HTML]{000000} \textit{Clarity}}              & {\color[HTML]{000000} whether the image is clear and its details are prominent.}                                                         \\
\cellcolor[HTML]{FFFFFF}{\color[HTML]{000000} }                                 & {\color[HTML]{000000} \textit{Visual Balance}}       & {\color[HTML]{000000} whether the visual elements in the image are evenly distributed.}                                                  \\
\rowcolor[HTML]{FFFFFF} 
\multirow{-8}{*}{\cellcolor[HTML]{FFFFFF}{\color[HTML]{000000} \textbf{Image}}} & {\color[HTML]{000000} \textit{Focus of the Picture}} & {\color[HTML]{000000} whether the image focuses on the subject, avoiding visual dispersion.}                                             \\ \hline
\end{tabular}}
\caption{Action space for text and image content generation.}
\label{tab:action-space}
\end{table*}

\underline{\textit{Training Data.}} We construct the training data not rely on explicit social network topology by assuming a fully connected graph over the audiences of each content.
We formulate each training sample as a quadruple \((u_i, u_j, c, y_{ijc})\), where \(u_i\) and \(u_j\) are user pairs, \(c\) is the content, and \(y_{ijc}\in\{0,1\}\) is the label showcasing the interaction behavior.  For a given content \(c\), let \(u_o\) denote the original poster and \(U_c\) the set of users who interacted with \(u_o\) (e.g., via retweet or reply). A sample is labeled as positive (\(y_{ijc} = 1\)) if either (1) \(u_i = u_o\) and \(u_j \in U_c\), or (2) both \(u_i, u_j \in U_c\). All other user pairs are labeled as negative (\(y_{ijc} = 0\)), indicating no observed interaction related to \(c\) between \(u_i\) and \(u_j\).
To mitigate class imbalance, we perform negative sampling by randomly dropping a subset of training samples with \(y_{ijc} = 0\), ensuring that the number of negative samples approximately matches the number of positive ones for each content.

\underline{\textit{Loss Function.}}
To train the pairwise diffusion estimator, we minimize a binary cross-entropy loss, defined as:
\begin{equation}
    \mathcal{L}_{CE} = \frac{1}{M} \sum_{i,j,c} \left[ y_{ijc} \log p_{ij}(c) + (1 - y_{ijc}) \log \left(1 - p_{ij}(c)\right) \right]
\end{equation}
where \(M\) is the total number of training instances, \(p_{ij}(c)\) is the output of the predicted diffusion estimator, and \(y_{ijc} \in \{0, 1\}\) is the groundtruth label.
The first term penalizes the model when it assigns a low probability to observed diffusion events, encouraging high confidence in positive cases. The second term penalizes false positives, pushing the model to assign low scores when diffusion does not occur. 

\begin{figure}
    \centering
    \includegraphics[width=1.0\linewidth]{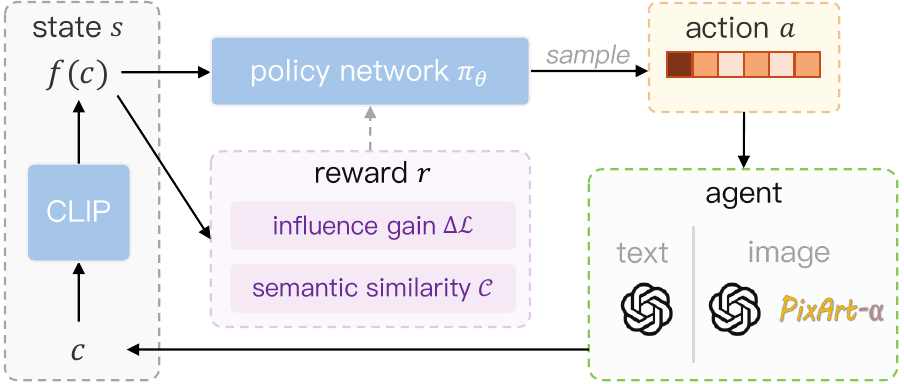}
    \caption{Our information editor for editing message content $c$. The policy network generates editing actions to iteratively revise the content, guided by the reward signal.}
    \label{fig:rl-pipeline}
\end{figure}

\subsection{Information Editor}

Figure~\ref{fig:rl-pipeline} illustrates the algorithm workflow of the information editor introduced in our framework. It rewrites the text or image content \(c\) to improve its diffusion impact. The content \(c\), along with its CLIP features \(f(c)\), is treated as the \emph{state} \(s\). Based on the state, the policy network \(\pi_\theta\), parameterized by \(\theta\), produces a conditional distribution \(\pi_\theta(a | s)\) over editing actions and samples an interpretable \emph{action} vector \(a\). This action is then passed to an \emph{agent} (i.e., GPT-4 for text revision or GPT-4 combined with PixArt-$\alpha$ for image generation), which produces the revised content representing the next state.
To train the policy network and guide the rewriting process toward maximizing diffusion impact, a \emph{reward} is computed by combining two parts: (i) {influence gain} $\Delta\mathcal{L}$, quantifying the improvement of the influence score as estimated by the influence indicator, and (ii) {semantic similarity} $\mathcal{C}$, measuring semantic consistency of the rewritten content and the original message. We describe each of these algorithm details in the rest of the section.

\underline{\textit{Action.}}  
We define an interpretable action vector \( a \), where each dimension \( a_i \in [-1, 1] \) (as detailed in Table~\ref{tab:action-space}) corresponds to a specific content feature to be modified. The value of \( a_i \) indicates the degree and direction of modification: \( a_i \approx +1 \) enhances the \( i \)-th feature, \( a_i \approx 0 \) leaves it unchanged, and \( a_i \approx -1 \) suppresses it. For text content, we adopt the STEPPS framework~\cite{pressgrove2018contagious}, which includes six dimensions—\textit{Social Currency}, \textit{Triggers}, \textit{Emotion}, \textit{Public}, \textit{Practical Value}, and \textit{Stories}—each empirically linked to increased virality. For image content, the action space comprises eight perceptual dimensions related to visual engagement, such as \textit{Colorfulness}, \textit{Brightness}, and \textit{Visual Balance}. These dimensions are informed by Li and Xie~\cite{li2020picture}, who identified image characteristics (e.g., color variation, emotional expression, and photographic quality) as key drivers of user engagement on social media.

\underline{\textit{Agent.}} 
The agent leverages prompt engineering to rewrite content using large language models, such as GPT-4 in our implementation (implementation details are available in the Appendix). Briefly, we embed the editing \textit{action} into the prompt, guiding the model to perform controlled rewriting.
For text content, we construct a dynamic prompt that encodes the action vector values (ranging from $-100\%$ to $100\%$) across six STEPPS dimensions. Each dimension is associated with a natural language instruction, annotated with the action vector value.
For image content, we adopt a similar strategy to construct the text-to-image prompt, guided by eight predefined visual dimensions. The resulting prompt is then input into the image generation model (e.g., PixArt-\(\alpha\) in our implementation), enabling controlled manipulation of visual features that affect diffusion.

\underline{\textit{Reward.}}
The reward quantifies the quality of a revision by jointly considering diffusion improvement and semantic fidelity. It is defined as:
\begin{equation}\label{eq:reward}
  r =
\begin{cases}
\sqrt{\Delta\mathcal{L} \cdot \mathcal{C}} & \Delta\mathcal{L} \ge 0 \\[6pt]
-\sqrt{-\Delta\mathcal{L} \cdot (1 - \mathcal{C})} & \Delta\mathcal{L} < 0
\end{cases}
\end{equation}
where the influence gain \(\Delta \mathcal{L}\) and the semantic similarity \(\mathcal{C}\) are defined as:
\begin{align}
\Delta \mathcal{L} &= \mathcal{L}_U(c) - \mathcal{L}_U(c^{\text{ori}}) \label{eq:deltaL} \\
\mathcal{C} &= \cos(f(c), f(c^{\text{ori}})) \label{eq:similarity}
\end{align}
where \(c^{\text{ori}}\) is the original input message content, \(\cos\) denotes the cosine similarity.
This formulation rewards the edits that boost diffusion while preserving semantic fidelity, and penalizes changes that are either semantically inconsistent or reduce diffusion potential.

\underline{\textit{Policy Network.}}  
The policy network defines a conditional distribution \(\pi_\theta(a | s)\) over editing actions \(a\), given the current state \(s\), represented by the content feature vector \(f(c)\). To generate a continuous action vector \(a \in [-1, 1]\), we use a reparameterization strategy:
\begin{equation}
    a = \tanh\left(\mu_\theta(c) + \sigma_\theta(c) \cdot \epsilon\right)
\end{equation}
where \(\epsilon \sim \mathcal{N}(0, I)\) is standard Gaussian noise that introduces controlled stochasticity to encourage exploration. The hyperbolic tangent function \(\tanh(\cdot)\) bounds the output within \([-1, 1]\), promoting numerical stability and interpretability. The mean \(\mu_\theta(c)\) and standard deviation \(\sigma_\theta(c)\) are predicted by a two-layer MLP applied to \(f(c)\), enabling the policy to flexibly capture uncertainty in the action space.

The objective is to learn an optimal policy that maximizes the expected reward:
\begin{equation}
    \pi^*(a | s) = \arg\max_{\pi_\theta} \sum_{a} \pi_\theta(a | s) \cdot Q(s,a)
\end{equation}
where \(Q(s, a)\) is the expected reward for applying action \(a\) to state \(s\), and \(\pi_\theta\) is the policy parameterized by \(\theta\).
We optimize the policy network using the policy gradient method~\cite{sutton1999policy}, which updates parameters to maximize the expected cumulative reward:
\begin{equation}
    \theta \leftarrow \theta + lr \cdot \frac{1}{B} \sum_{i=1}^B \sum_{t=1}^T \nabla_\theta \log \pi_\theta(a_i^{(t)} | s_i^{(t)}) \cdot J_{i,t}
\end{equation}
where \(i\) indexes the sample in the batch, \(a_i^{(t)}\) is the action taken at step \(t\), \(s_i^{(t)}\) is the state at step \(t\), \(\nabla_\theta\) is the gradient with respect to policy parameters \(\theta\), \(B\) is the batch size, \(\gamma\) is the discount factor, \(lr\) is the learning rate, and \(J_{i,t}\) is the discounted return from step \(t\), defined as:
\begin{equation}
    J_{i,t} = \sum_{k=t}^T \gamma^{k-t} r_{i,k}
\end{equation}
in which \(r_{i,k}\) is the reward at step \(k\) computed by Eq.~(\ref{eq:reward}).

\begin{table}[t]
\centering
\setlength{\tabcolsep}{1mm}
{
\begin{tabular}{l|c|c|c|c}
\toprule
{Dataset} & {Seed Tweets} & {Users} & {All Tweets} & {Time Period} \\
\midrule
Olympics &4,375& 10,097 & 5,812,520 & 2024.8-2024.11 \\
Movie &6,625& 14,552 & 6,181,779 & 2022.4-2024.11 \\
\bottomrule
\end{tabular}
}
\caption{Statistics of datasets.}
\label{tab:datasets}
\end{table}

\begin{figure}[t]
  \centering
  \includegraphics[height=3.6cm]
  {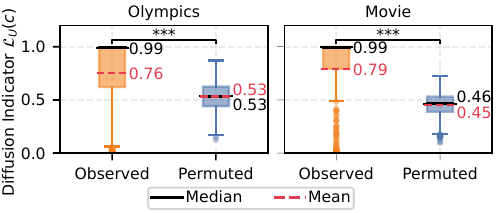}
  \caption{Distribution of influence scores for messages on \emph{observed}~(orange) and \emph{permuted}~(blue) combinations. *** denotes statistical significance at $p < 0.001$.}
  \label{fig:diffusion_boxplots}
\end{figure}

\begin{table}[t]
\centering
\setlength{\tabcolsep}{1mm}
{
\begin{tabular}{l|c|c|c|c}
\toprule
{Dataset} & {Precision} & {Recall} & {F1-score} & {AUC} \\
\midrule
Olympics &0.8187 &0.8122 &0.8113 &0.9042 \\
Movie &0.8360 &0.8286 &0.8277 &0.9186  \\
\bottomrule
\end{tabular}
}
\caption{Performance of the pairwise diffusion estimator.}
\label{tab:metrics}
\end{table}

\begin{figure*}[t]
    \centering
    \includegraphics[width=1.0\linewidth]{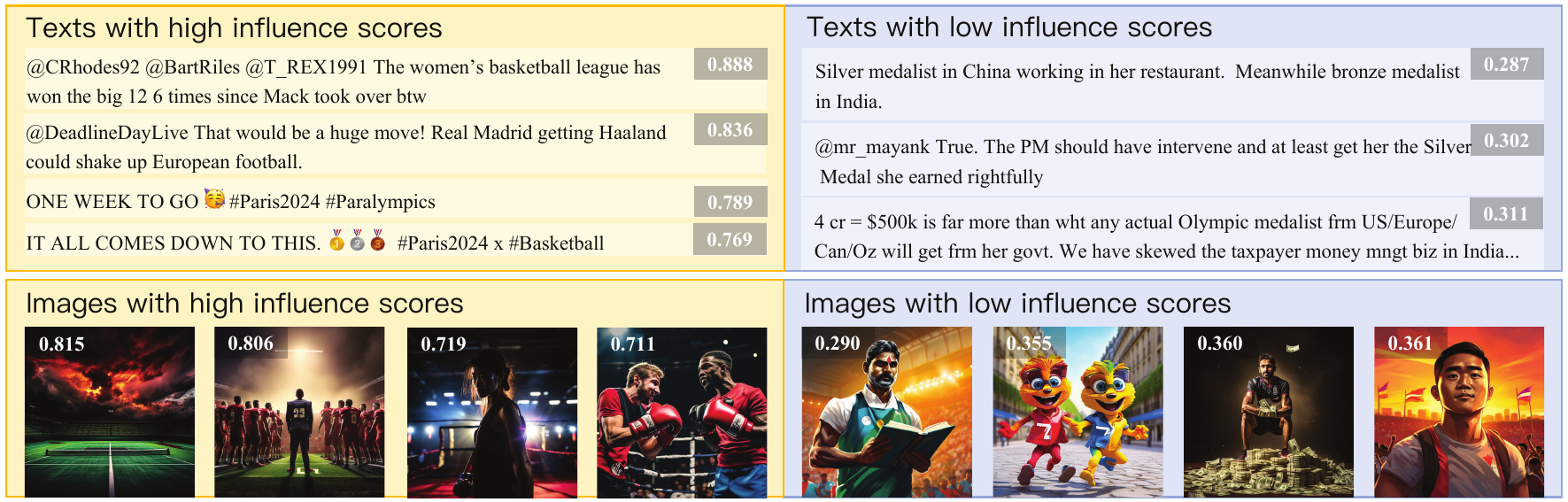}
\caption{Representative text and image examples across different levels of influence scores \(\mathcal{L}_U(c)\).}
    \label{fig:lc_examples}
\end{figure*}

\section{Evaluation}
We conducted comprehensive experiments to evaluate the effectiveness of our proposed framework. In this section, we begin by describing the dataset details. We then separately evaluate the influence indicator and the information editor. Finally, we present results from a user study designed to assess whether the generated content aligns with human preferences in the real-world information diffusion scenarios.

\subsection{Datasets}

We constructed two X (Twitter) datasets focused on discussions about recent movies and the Paris 2024 Summer Olympics (Table~\ref{tab:datasets}) using a three-step process. First, we collected tweets containing topic-relevant keywords, referred to as \emph{seed tweets}. Second, for each seed tweet, we defined its audience group as the original poster and users who directly interacted with it via retweets or replies. Third, we merged all audience groups to form a unified topic-specific audience group, and collected their tweets within a defined time window to build the final dataset. In total, the datasets comprise over 10{,}000 users and 11 million tweets.


\subsection{Indicator Estimation}
In this section, we present the implementation and experimental setup of the influence indicator, followed by an evaluation of its effectiveness from two perspectives:  
(1) the predictive capacity of the pairwise diffusion estimator  $p_{ij}(c)$, and  
(2) the ability of the influence score \(\mathcal{L}_U(c)\) to capture variations in diffusion impact across different audience groups.  
We also provide qualitative examples of content with high and low influence scores to demonstrate the practical utility of the proposed indicator.

\underline{\textit{Implementation \& Settings.}}  
We trained the pairwise diffusion estimator using the Adam optimizer with a learning rate of $1 \times 10^{-4}$. Input features were standardized by removing the mean and scaling to unit variance along each dimension. The dataset was split into training and test sets in a 4:1 ratio, with no overlap in users or messages between the two sets.

\underline{\textit{Results.}}  
As shown in Table~\ref{tab:metrics}, the pairwise diffusion estimator demonstrates strong predictive performance, achieving AUC scores of 0.9042 on {Olympics} and 0.9186 on {Movie}, indicating its effectiveness in distinguishing diffusion from non-diffusion behaviors.
To evaluate the influence score \(\mathcal{L}_U(c)\), we compare its values on \emph{observed} message–audience combinations \((c_i, U_i)\) against \emph{permuted} combinations \((c_i, U_j)\) with \(i \neq j\),  where \(U_i\) is the actual audience group for message \(c_i\) and \(U_j\) is unrelated. Figure~\ref{fig:diffusion_boxplots} shows that observed combinations yield significantly higher scores (mean: 0.76 vs.\ 0.53 for {Olympics}; 0.79 vs.\ 0.45 for {Movie}), with minimal overlap between distributions. A Mann–Whitney \(U\) test confirms these differences are highly significant (\(p < 0.001\)), demonstrating that \(\mathcal{L}_U(c)\) effectively captures audience-specific diffusion potential.

\underline{\textit{Visualization.}}  
Figure~\ref{fig:lc_examples} presents representative messages from the {Olympics} dataset with low and high influence scores. Low-scoring examples typically lack topical relevance or suffer from weak linguistic or visual quality. In contrast, high-scoring messages exhibit strong contextual alignment, coherent composition, and features~(e.g., emotion) that are more likely to engage the target audience.

\subsection{Editor Estimation}
In this section, we firstly describe the implementation details and experimental setup of information editor. We then outline the baselines and evaluation metrics, and present both quantitative and qualitative comparisons with baselines. Finally, we conduct an ablation study on the reward function.

\underline{\textit{Implementation \& Settings.}}  
We trained the policy network for 350 episodes with a trajectory length of 3, using the Adam optimizer with a learning rate of $1 \times 10^{-4}$. For image content generation, each input message was first converted into a text-to-image prompt by GPT-4 and then processed by PixArt-$\alpha$ to generate the initial image.
For evaluation, we sampled 200 messages from the {Olympics} test set and computed their influence scores. The 40 messages with the lowest scores were used as the test set, while the remaining 160 messages formed the training set.

\underline{\textit{Baselines.}}  
We compare our framework with the following baselines (detailed implementation could be found in the appendix):
(1)~\textit{Large Language Model (LLM)}: directly prompts a pretrained language model to rewrite the input message, without any audience-specific guidance;
(2)~\textit{In-Context Learning (IC-L)}: for each audience group, retrieves messages with the highest and lowest observed influence scores as positive and negative exemplars, respectively, and uses them as few-shot prompts to guide LLM-based rewriting;
(3)~\textit{Greedy Search (Greedy)}: as a non-learning variant of our framework, uniformly samples editing actions and selects the revised message with the highest reward at each step, without learning a policy network.

\underline{\textit{Metrics.}}  
We assess our information editor using two metrics:  
(1) \emph{Diffusion Gain}, which measures the relative increase of influence scores after rewriting, and 
(2) \emph{Consistency}, which quantifies semantic preservation between original and edited content via the cosine similarity of their features.  The details can be found in the appendix.

\begin{table}[h]
\centering
\setlength{\tabcolsep}{1mm}
{
\begin{tabular}{lc|c|c}
\hline
 & & \multicolumn{1}{c|}{Diffusion Gain$\uparrow$} & \multicolumn{1}{c}{Consistency $\uparrow$} \\ \hline
\multicolumn{1}{l|}{\multirow{4}{*}{Text}}  & \textbf{Ours} & \textbf{20.69\%} & 0.9510        \\
                       \multicolumn{1}{l|}{}  & LLM           & 14.03\%         & \textbf{0.9590} \\
                       \multicolumn{1}{l|}{}  & IC-L          & 13.43\%         & 0.9509         \\
                       \multicolumn{1}{l|}{}  & Greedy        & 17.93\%         & 0.9470         \\ \hline
\multicolumn{1}{l|}{\multirow{4}{*}{Image}} & \textbf{Ours} & \textbf{11.20\%} & 0.8730         \\
                      \multicolumn{1}{l|}{}   & LLM           & 1.36\%          & 0.8743         \\
                      \multicolumn{1}{l|}{}   & IC-L          & 3.05\%          & \textbf{0.8748} \\
                      \multicolumn{1}{l|}{}   & Greedy        & 7.93\%          & 0.8693         \\ \hline
\end{tabular}
}
\caption{Quantitative comparison with baselines.}
\label{tab:quantitative_comparison}
\end{table}

\begin{figure*}[tbp]
    \centering
    \includegraphics[width=1.0\linewidth]{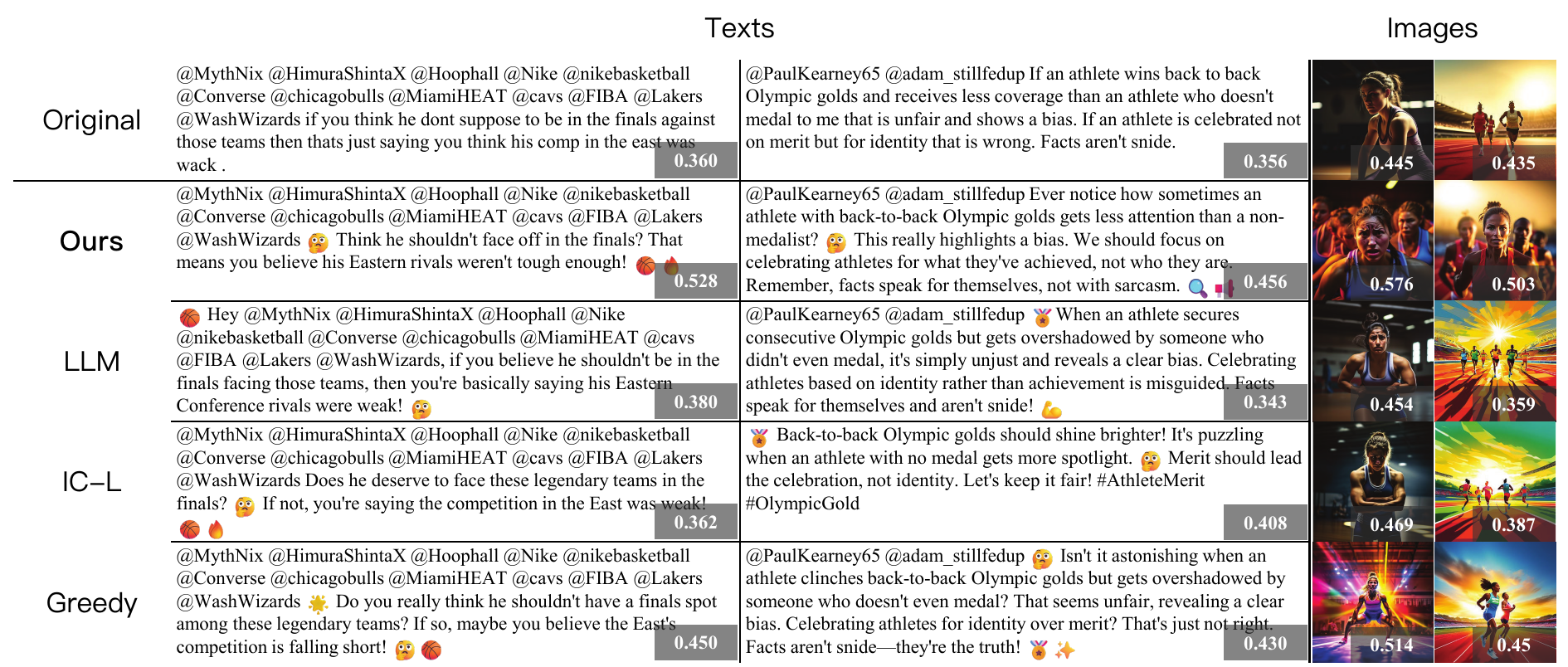}
    \caption{Qualitative comparison with baselines. Each piece of content is annotated with its corresponding influence score.}
    \label{fig:qualitative_comparison}
\end{figure*}

\underline{\textit{Quantitative Comparison.}}  
Table~\ref{tab:quantitative_comparison} presents the performance on both text and image generation tasks. In the text task, our method achieves the highest Diffusion Gain at 20.69\%, significantly outperforming other baselines. Additionally, our method attains a high Consistency score (0.9510), indicating effective semantic preservation. Although LLM achieves slightly higher Consistency (0.9590), its diffusion improvement is considerably weaker.
For the image task, our method again leads with a Diffusion Gain of 11.20\%, surpassing other baselines. In terms of Consistency, IC-L (0.8748) and LLM (0.8743) slightly exceed our score (0.8730). However, the difference is minimal, confirming that our edits maintain semantic integrity while delivering significantly better diffusion performance.

\underline{\textit{Qualitative Comparison.}}  
Figure~\ref{fig:qualitative_comparison} illustrates representative generation results for both text and image content. In the text examples, our framework transforms generic statements into vivid, action-oriented language that heightens emotional engagement and topical relevance while preserving the original intent. In the image examples, our approach enhances visual appeal by emphasizing human subjects, resulting in more compelling and audience-aware imagery.

\begin{table}[t]
    \centering
    \setlength{\tabcolsep}{1mm}
    {
        \begin{tabular}{lc|c|c}
            \hline
            & & Diffusion Gain$\uparrow$ & Consistency $\uparrow$ \\ \hline
            \multicolumn{1}{l|}{\multirow{2}{*}{Text}} 
                & $\sqrt{\Delta\mathcal{L}\cdot \mathcal{C}}$ & 20.69\% & \textbf{0.9510} \\
            \multicolumn{1}{l|}{} 
                & $\Delta \mathcal{L}$ & \textbf{21.98\%} & 0.9471 \\ \hline
            \multicolumn{1}{l|}{\multirow{2}{*}{Image}} 
                & $\sqrt{\Delta\mathcal{L}\cdot \mathcal{C}}$ & 11.20\% & \textbf{0.8730} \\
            \multicolumn{1}{l|}{} 
                & $\Delta \mathcal{L}$ & \textbf{12.43\%} & 0.8707 \\ \hline
        \end{tabular}
    }
    \caption{Results of ablation study.}
    \label{tab:ablation}
\end{table}

\underline{\textit{Ablation Study.}}  
We assess the reward function by comparing the variant optimizing only the influence gain component \(\Delta\mathcal{L}\). Table~\ref{tab:ablation} shows the results. When trained with only the influence gain, the model boosts diffusion but sacrifices semantic accuracy. This highlights the importance of incorporating semantic similarity, as it ensures that the content remains meaningful while still achieving improved diffusion.

\subsection{User Study}

We conducted a user study to evaluate whether the generated content increases users’ likelihood of sharing information in realistic settings.

\underline{\textit{Procedure \& Metric.}}  
We conducted a user study on the Prolific platform~\cite{prolific}, recruiting native English speakers with informed consent. Participants were screened based on their X activity and recent engagement with the target topic. A total of 100 qualified users were selected (76 male, 24 female; ages: 18--25 (11\%), 26--50 (68\%), 51--60 (21\%)). Each participant evaluated 100 tweet pairs and 100 image pairs, each consisting of an original and a revised version of either text or visual content. For each pair, participants selected the version they were more likely to share (via retweet or reply). To mitigate ordering bias, all pairs were presented in randomized order.

We evaluated performance using the \textbf{\textit{Retweet Preference Rate} (RPR)}—the average proportion of times the rewritten version was preferred by our participants.

\underline{\textit{Results \& Analysis.}}  
The results indicate that the generated content significantly increased participants’ willingness to retweet. The overall RPR was \textbf{62.36\%}(\textbf{64.00\%} for text and \textbf{60.72\%} for image), suggesting that enhancements in both modalities contribute to improved diffusion impact. A paired-sample t-test confirmed that the overall preference for the rewritten content is statistically significant (\(p < 0.001\)).
We further analyzed participants’ stated reasons for their choices. Common reasons included stronger emotional resonance and alignment with personal experiences.  

\section{Discussion and Conclusion}

In this work, we introduce a novel task, \emph{Diffusion-Oriented Content Generation}, and propose a reinforcement learning (RL)–based framework that rewrites messages to enhance diffusion among targeted audiences. Central to our approach is an influence indicator that estimates the diffusion potential of content for a given audience, without relying on explicit network topology. Based on this indicator, we develop an information editor that optimizes message rewrites for maximal diffusion. We model the editor as a finite-horizon MDP and adopt a stationary policy for practicality, which remains effective even though optimal policies are generally non-stationary. Experiments on real-world datasets show that our indicator accurately predicts diffusion outcomes, and that the RL-based rewriting strategy yields significant diffusion gains while maintaining semantic fidelity.

Despite promising performance, our method still has limitations. Our current iterative rewriting strategy incurs significant latency per instance. Integrating diffusion feedback directly into the generative model could mitigate this, but real‑world diffusion signals are often noisy, hindering reliable feedback integration. Our method currently does not account for temporal variations in audience features, and incorporating such dynamics remains an important direction for future research. Another future work is to explore more modalities beyond text and image. Extending the framework to audio and video could enhance its practicality. 

\section{Acknowledgments}
Nan Cao is the corresponding author. This work was supported by the National Key Research and Development Program of China (2023YFB3107100).

\bibliography{aaai2026}

\end{document}